\documentclass[a4paper,11pt]{article}
\usepackage{pos}
\usepackage{amsmath}

\renewcommand{\logo}{\relax}

\usepackage{graphicx}
\usepackage{float}
\usepackage{subcaption}
\newcommand{\powhegVersion}{\textsc{POWHEG-BOX-V}2}

\newcommand{\openloopsVersion}{\textsc{OpenLoops}~2}

\newcommand{\pythiaVersion}{\textsc{Pythia}~8.317}

\newcommand{\rivetVersion}{\textsc{Rivet}~4}

\title{Top-Quark Pair Production and Decay in Extended Higgs Sectors}

\author[a]{Andrea Banfi}
\author[b]{Alexander Lind}
\author[a]{Jonas M. Lindert}
\author*[c]{Vishakha Lingadahally}

\affiliation[a]{Department of Physics and Astronomy, University of Sussex,\\
  Brighton BN1 9QH, United Kingdom}

\affiliation[b]{INFN, Sezione di Roma,\\ Piazzale Aldo Moro 5, 00185 Roma, Italy}

\affiliation[c]{Department of Physics, Royal Holloway, University of London,\\
Egham TW20 0EX, United Kingdom}

\emailAdd{a.banfi@sussex.ac.uk}
\emailAdd{alexander.lind@roma1.infn.it}
\emailAdd{j.lindert@sussex.ac.uk}
\emailAdd{vishakha.lingadahally.2023@live.rhul.ac.uk}

\abstract{Extensions of the Higgs sector in the Standard Model address outstanding questions like the first-order electroweak phase transition and also present a rich collider phenomenology. The two-Higgs-doublet model is a minimal extension of the Standard Model with an extra complex $SU(2)_L$ doublet, and consists of an extra scalar, a pseudoscalar and a pair of charged Higgs bosons along with a Standard-Model-like Higgs boson. We present a modified version of the \texttt{hvq} library of the \powhegVersion{} event generator interfaced with the \openloopsVersion{} amplitude generator to obtain predictions for top-quark pair production at NLO accuracy, including top-quark decays with full spin correlations in a framework that extends the SM Higgs sector with an extra heavy scalar and a pseudoscalar. The interference effects between the scalar and pseudoscalar signals, and the Standard Model QCD background are implemented. We include parton showers and hadronisation, and take detector effects into consideration. The status of our work and plans to extend it to future collider projects are discussed.}

\FullConference{43rd International Conference on High Energy Physics (ICHEP 2026)\\
30 July  to 5 August , 2026\\
Natal, Brazil\\}

\begin{document}
\maketitle

\section{Introduction}

Extensions of the Standard Model Higgs sector containing additional scalars and pseudoscalars arise in a wide variety of Beyond-the-Standard-Model (BSM) scenarios and have a rich collider and cosmological phenomenology. For example, the two-Higgs-doublet model (2HDM) extends the scalar sector by a second complex $SU(2)_L$ doublet and consists of two CP-even scalars, a CP-odd pseudoscalar and a pair of charged Higgs bosons~\cite{Branco_2012}.

Heavy neutral Higgs bosons with masses above the top quark pair production threshold can be probed through their decays to $t\bar{t}$~\cite{Kauer_2019,Banfi:2023zav}. The large top-Yukawa coupling makes this final state particularly sensitive to additional scalar and pseudoscalar states that couple appreciably to the top quark. The process of top-quark pair production through the decay of heavy scalars and pseudoscalars is associated with large interference effects between the scalar and pseudoscalar signals, and the QCD background. Such effects have been studied for top-quark pair production through the decay of a heavy scalar in the one-Higgs-singlet model (1HSM) at NLO accuracy recently~\cite{Banfi:2023zav}. Signal--background interference is also explicitly incorporated in current ATLAS and CMS searches for additional scalar and pseudoscalar states in the $t\bar{t}$ channel at LO accuracy~\cite{2024,Collaboration_2025}. A reliable description of this process therefore requires the signal and interference contributions to be treated consistently.

We perform computations of the cross section for top-quark pair production at NLO accuracy, including leptonic top-quark decays with full spin correlations in a framework that extends the SM Higgs sector with an extra heavy scalar and a pseudoscalar. In particular, we include the large interference effects between the scalar and pseudoscalar signals, and the QCD background. The interference can be comparable to, or larger than, the pure signal contribution and can transform the conventional resonant enhancement into a characteristic peak--dip or predominantly dip-like structure.  We implement our work in a modified version of the \texttt{hvq} library~\cite{Frixione:2007nw,Frixione:2007nu} in \powhegVersion{}~\cite{Nason:2004rx,Frixione:2007vw,Alioli:2010xd} interfaced with \openloopsVersion{}~\cite{Buccioni_2019,Ossola:2007ax,vanHameren:2010cp}.

\section{Structure of the NLO QCD Computation}

The total squared amplitude for the process $pp{\text{ }}(\to h/H/A_0)\to t\overline{t}+X$, where $h$ is the Standard-Model-like Higgs boson, $H$ is the CP-even heavy Higgs boson and $A_0$ is the CP-odd Higgs boson, can be expressed as follows:
\begin{equation}
    \left \vert \mathcal{M} \right \vert^2 \approx \left \vert \mathcal{M}_{\rm QCD} \right \vert^2 + \left \vert \mathcal{M}_{h} \right \vert^2 + \left \vert \mathcal{M}_{H} \right \vert^2 + \left \vert \mathcal{M}_{A_0} \right \vert^2 + 2 \, {\rm Re} \left ( \mathcal{M}_{\rm QCD}^{\ast} \left ( \mathcal{M}_{h} + \mathcal{M}_{H} + \mathcal{M}_{A_0} \right )  \right ) \,,
\end{equation}

where $\mathcal{M}_{\rm QCD}$, $\mathcal{M}_{h}$, $\mathcal{M}_{H}$ and $\mathcal{M}_{A_0}$ are the QCD background, the SM-like Higgs boson, the CP-even heavy Higgs boson and the CP-odd Higgs boson amplitudes respectively. We neglect the interference contributions arising from the interference between the SM-like Higgs boson $h$ and the CP-even heavy Higgs boson $H$. The interference between the CP-odd Higgs boson $A_0$ and CP-even Higgs boson contributions cancel out to zero. The term $2\,{\rm Re}\big(\mathcal{M}_{\rm QCD}^*(\mathcal{M}_{h}+\mathcal{M}_H+\mathcal{M}_{A_0}) \big)$ is the interference between the scalar and pseudoscalar amplitudes, and the QCD background amplitude. 

The one- and two-loop form factors for the scalar and pseudoscalar Higgs production are implemented in a novel form factor interface in \openloopsVersion{} as an extension to the one introduced in ref.~\cite{Banfi:2023zav}. The code for the form factors is adapted from the implementations in \textsc{JetVHeto}~\cite{Banfi:2012jm,Banfi:2016} and the \texttt{gg\_H\_quark-mass-effects} and \texttt{gg\_H\_2HDM} processes~\cite{Bagnaschi:2011tu} in \powhegVersion{}. Following the procedure employed in 1HSM at NLO in ref.~\cite{Banfi:2023zav}, the non-factorisable corrections to the interference at NLO are implemented in the heavy-top limit. The one-loop non-factorisable corrections in the heavy-top limit are rescaled with the exact Born amplitude. This approximation of the non-factorisable corrections is valid in the soft limit, and allows for the cancellation of infrared divergences between the virtual and real correction contributions. 

We study top-pair production with subsequent leptonic $W$-decay,
\begin{equation}
    pp \, \left ( \to h/H/A_0 \right ) \to t\bar{t} + X \to b\bar{b} e^{\pm} \mu^{\mp} \nu \bar{\nu} + X \,.
\end{equation}

The methodology of preserving spin correlation effects in the decay products implemented in the \texttt{hvq} library of \powhegVersion{} is described in detail in ref.~\cite{Alioli:2011as}. The squared production matrix elements are reweighted with the squared decay matrix elements to construct the decay phase space where the spin correlations in the decay products are retained through a `hit-and-miss' procedure. Parton showering and hadronisation are implemented after the top-quark decays with \pythiaVersion~\cite{bierlich2022comprehensiveguidephysicsusage} and the analysis of the results is performed with \rivetVersion~\cite{bierlich2024robustindependentvalidationexperiment}. We perform
particle-level reconstruction and apply fiducial cuts on leptons and
$b$-tagged jets.

As per the recommendations of the LHC Higgs Working Group (LHCHWG)~\cite{LHCHiggsCrossSectionWorkingGroup:2016ypw}, we use the $G_\mu$ scheme with input parameters from the latest PDG values~\cite{doi:10.1142/S0217751X26300115} for masses and widths of the top quark, the SM-like Higgs boson, the $Z$ boson, and the $W$ boson. We neglect the masses of the five lighter quark flavours and the leptons. We use the \texttt{PDF4LHC15\_nlo\_mc}~\cite{Butterworth_2016} PDF set for our predictions in Run 2.

The dynamic renormalisation and factorisation scales $\mu_R$ and $\mu_F$ are chosen to be
\begin{equation}
    \mu_R = \mu_F = \frac{M_{t\bar{t}}}{2} \,,
\end{equation}
where $M_{t\bar{t}}$ is the invariant mass of the top-quark pair evaluated at the Monte Carlo truth-level. 

\section{Preliminary Results}

We obtain differential cross sections as functions of spin-sensitive observables with the SM-like Higgs boson as well as an SM-like CP-odd Higgs boson for Run 2. In case of the CP-odd Higgs boson, we set the mass of the pseudoscalar $m_A$ to 500 GeV, and the width of the pseudoscalar $\Gamma_A$ to 21.3 GeV which correspond to the values of $m_A$ and $\Gamma_A$ used in ref.~\cite{Hespel_2016}. It is referred to as an `SM-like CP-odd Higgs boson' or an `SM-like pseudoscalar', because the coupling of the pseudoscalar to the top quark is set to that of the SM Yukawa value. Our results for the inclusive cross section for the top-quark pair produced by the decay of the SM-like Higgs boson agree with those presented in ref.~\cite{Banfi:2023zav} at LO and NLO accuracy, and ref.~\cite{Hespel_2016} at LO accuracy. Our results for the inclusive cross section for the top-quark pair produced by the decay of the SM-like pseudoscalar at LO accuracy agree with those presented in ref.~\cite{Hespel_2016}.

\begin{figure}[H]
    \centering
    \begin{subfigure}[c]{0.48\textwidth}
        \centering
    \includegraphics[width=\linewidth]{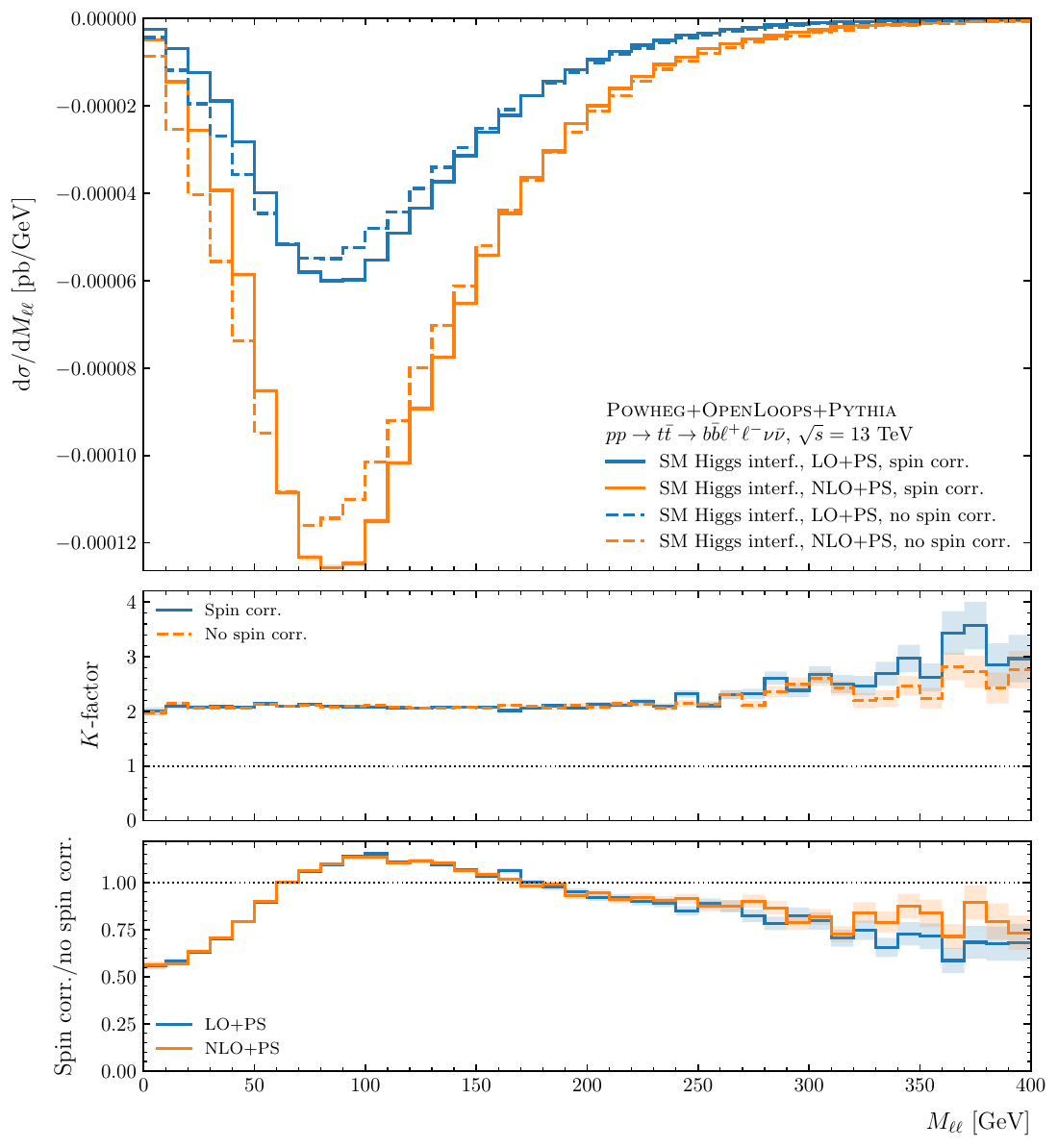}
        \caption*{}
        \label{H-interf}
    \end{subfigure}
    \hfill
    \begin{subfigure}[c]{0.48\textwidth}
        \centering
        \includegraphics[width=\linewidth]{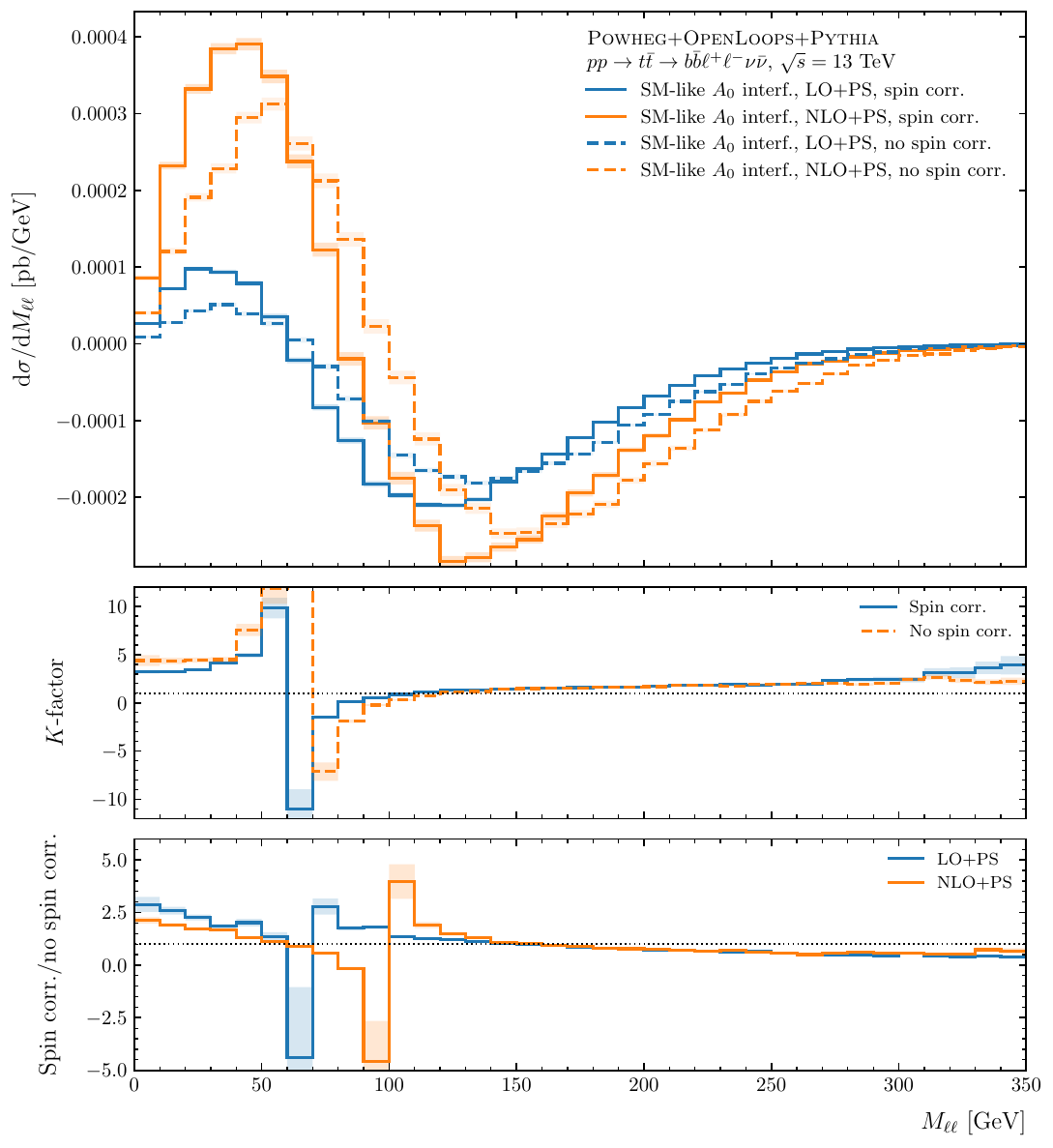}
        \caption*{}
        \label{A0-interf}
    \end{subfigure}
    \captionsetup{skip=0.5pt}
  \caption{Differential cross sections for the interference between the scalar (left) and pseudoscalar (right) signals and the QCD background in top-quark pair production and subsequent leptonic decays, as functions of the dilepton invariant mass $M_{\ell\ell}$, at LO+PS and NLO+PS accuracy.}
    \label{M_ll}
\end{figure}

In the differential cross sections as functions of the invariant mass of the outgoing lepton pair in fig.~\ref{M_ll}, we notice a marked difference in shapes of the distributions when the spin correlation effects are considered and when they are not. In case of the interference between the pseudoscalar signal and the QCD background (right panel of fig.~\ref{M_ll}), the peak--dip structures of the differential cross sections as functions of the invariant mass of the leptons at LO+PS and NLO+PS show a noticeable difference in shape from one another as well. The interference between the pseudoscalar signal and the QCD background consists of both positive and negative contributions. Consequently, the inclusive $K$-factor which is less than unity in this case, does not fully capture the effect of the higher-order corrections to the process. Therefore, it is instructive to examine the cross section at the differential level where the nuanced peak--dip structure is revealed. The NLO+PS prediction has an enhanced peak and a deeper dip compared to LO+PS, demonstrating that higher-order corrections accentuate the structure of the interference despite the smaller inclusive $K$-factor.

\section{Outlook}

With the novel tool we have developed to compute the cross section for top-quark pair production at NLO, including leptopnic decays with a heavy scalar and a pseudoscalar, we have produced preliminary results in a simplified extension of the SM Higgs sector and validated our results. We are working on performing phenomenological predictions with relevant benchmark points that are of high experimental interest in extensions of the SM Higgs sector, such as the 2HDM. The consistent implementation of interference effects between the scalar/pseudoscalar signals and the QCD background, as well as the full spin correlation effects in the top-quark decays, makes our tool very useful for experimental searches. The tool enables the study of higher-order effects on the shapes of the differential cross sections, especially the nuanced peak--dip interference structure. The inclusion of spin correlations in the decays allows us to probe the effects of spin correlation in the process through the study of differential cross sections as functions of spin-sensitive observables. We are working on extending the current numerical setup to that of Run 3 and HL-LHC, and studying the feasibility of detection of heavy scalars and pseudoscalars in future collider projects.

\acknowledgments

I would like to thank the INFN-LNF Theory Group for their hospitality during my visit. I am thankful to the organisers of ICHEP 2026 for the opportunity to present this work at the conference. This research is funded by the Science and Technology Facilities Council (STFC) of the UK Research and Innovation (UKRI) under Project 2888853.



\bibliographystyle{JHEP}
\bibliography{References}


\end{document}